# Tunable and switchable multi-wavelength erbium-doped fiber ring laser based on a modified dual-pass Mach-Zehnder interferometer

Ai-Ping Luo,[*] Zhi-Chao Luo,[**] and Wen-Cheng Xu

*Key Laboratory of Photonic Information Technology of Guangdong Higher Education Institutes, School of Information and Optoelectronic Science and Engineering, South China Normal University, Guangzhou, Guangdong 510006, China*

[*]Email: luoaiping@scnu.edu.cn

[**]Email: zcluo@scnu.edu.cn

A tunable and switchable multi-wavelength erbium-doped fiber ring laser based on a new type tunable comb filter is proposed and demonstrated. By adjusting the polarization controllers, dual-function operation of the channel spacing tunability and the wavelength switching (interleaving) can be readily achieved. Up to 29 stable lasing lines with 0.4 nm spacing and 14 lasing wavelengths with 0.8 nm spacing in 3 dB bandwidth were obtained at room temperature. In addition, the lasing output, including the number of the lasing lines, the lasing evenness and the lasing locations, can also be flexibly adjusted through the wavelength-dependent polarization rotation mechanism.

Multi-wavelength fiber lasers have attracted considerable attention for their potential applications in the fields of wavelength-division-multiplexing (WDM) transmission system, optical fiber sensing, fiber devices testing and measuring. They can be achieved by combining various kinds of optical filters. In some practical applications, it is also desirable to be able to enhance the functionality and flexibility of a multi-wavelength fiber laser. Therefore, the tunability of the wavelength spacing, the number of channels and the lasing locations should be investigated. Up to date, various techniques have been used to realize channel-spacing tunable operation of multi-wavelength lasing, such as using cascaded multisection polarization maintain fiber (PMF) [1], incorporating high birefringence fiber in the fiber loop mirror [2], employing multiple fiber Bragg gratings in the cavity [3].

Generally, multi-wavelength fiber laser based on a standard Mach-Zehnder (M-Z) interferometer is nontunable [4] except incorporating an optical variable delay line (OVDL) to change the path difference between two interferometer arms [5]. Nevertheless, the accurate tuning of the OVDL must occur through a controlling computer. In this letter, we propose and demonstrate a tunable and switchable multi-wavelength erbium-doped fiber ring laser by exploiting a new type comb filter based on a modified dual-pass M-Z interferometer. By adjusting the polarization controllers (PCs), multi-wavelength lasing with the channel spacing tunability between 0.4 nm and 0.8 nm as well as the wavelength switching (interleaving) can be easily achieved. Up to 29 stable lasing lines with 0.4 nm spacing and 14 lasing wavelengths with 0.8 nm spacing in 3 dB bandwidth were obtained at room temperature. Furthermore, the lasing output, including the number of the lasing lines，the lasing evenness and the lasing locations, can also be flexibly adjusted.

Fig. 1 shows the experimental setup of the proposed multi-wavelength erbium-doped fiber (EDF) laser. A 4.5 m long EDF serves as the gain medium. A 980 nm/1550 nm WDM is used to launch the 980 nm pumping laser into the laser cavity. Two PCs are employed to adjust the polarization states of the circulating light. A polarization-dependent isolator (PD-ISO) with fiber pigtails assures the unidirectional operation and provides the wavelength-dependent polarization rotation mechanism [6-7]. A phase modulator composed of a 4 m long single-mode fiber (SMF) wrapped around a cylindrical piezoelectric transducer (PZT) which has a resonant frequency of 23 kHZ is used to suppress the mode competition. The tunable comb filter, which is developed from a standard dual-pass M-Z interferometer, consists of a 50:50 and a 30:70 fiber couplers with a PC3 in one arm, as shown in Fig. 1. The laser output is measured by an optical spectrum analyzer (OSA).

First, the transmission characteristics of the individual modified dual-pass M-Z filter are analyzed. As the actual fiber is not ideal, birefringence varies randomly along the fiber. Consequently, for effectively realizing the tunable function, we employ a PC3 in one arm of the interferometer to control the influence of birefringence on the channel spacing. The initial polarization state of the input light is forced into aligning with the polarizer provided by the PD-ISO. Suppose that the linearly polarized light at an angle $\alpha$ with respect to one of the principal axes of the fiber is launched into the filter, where $\alpha$ can be adjusted by the PC1. The filter characteristics can be analyzed from the following Jones matrix representation:

$$\begin{bmatrix}[E_{1out}]\\[E_{2out}]\end{bmatrix}=[C_1]\begin{bmatrix}[P][F_1] & 0\\ 0 & [F_2]\end{bmatrix}[C_2]\begin{bmatrix}0 & [I]\\ [I] & 0\end{bmatrix}[C_2]\begin{bmatrix}[F_1][P] & 0\\ 0 & [F_2]\end{bmatrix}[C_1]\begin{bmatrix}[E_{1in}]\\[E_{2in}]\end{bmatrix} \quad (1)$$

In Eq.(1), $[E_{1in}]$ and $[E_{2in}]$ are the input fields of port 1 and port 2, respectively, $[E_{1in}]=[A\cos\alpha; A\sin\alpha]$ and $[E_{2in}]=[0;0]$, where $A$ is the amplitude of the light. $[I]$ is the identity matrix. $[C_m](m=1,2)$, $[P]$ and $[F_n](n=1,2)$ represent the matrices of the fiber couplers, the PC3 and two interferometer arms, respectively.

$$[C_m]=\begin{bmatrix}\sqrt{1-a_m}[I] & j\sqrt{a_m}[I] \\ j\sqrt{a_m}[I] & \sqrt{1-a_m}[I]\end{bmatrix}, \quad [P]=\begin{bmatrix}\cos\theta & -\sin\theta \\ \sin\theta & \cos\theta\end{bmatrix}, \quad [F_1]=\begin{bmatrix}e^{jkn_xL} & 0 \\ 0 & e^{jkn_yL}\end{bmatrix}, \quad [F_2]=\begin{bmatrix}e^{j(kn_xL+\varphi)} & 0 \\ 0 & e^{j(kn_yL+\varphi)}\end{bmatrix}.$$

where $a_m$ is the coupling ratio, $\theta$ is the rotation angle of the propagating light through the PC3, $L$ is the length of the shorter arm, $\varphi$ is the phase difference between the two arms which is determined by the path difference $\Delta L$ ($\Delta L = 2mm$ in this experiment). Then, we can obtain the transmission function in port 2:

$$\begin{aligned}T &= \left|\frac{E_{2out}}{E}\right|^2 \\ &= 2a_2(1-a_2)+(1-2a_2)^2\sin^2\theta\sin^2\frac{\delta}{2}+2a_2(1-a_2)[(\cos^2\theta-\sin^2\theta\cos\delta)\cos 2\varphi \\ &\quad +\sin\theta\sin(2\alpha+\theta)\sin\delta\sin 2\varphi]-2(1-2a_2)\sqrt{a_2(1-a_2)}\sin^2\theta\cos(2\alpha+\theta)\sin\delta\sin\varphi\end{aligned} \quad (2)$$

where $\delta = 2k(n_x - n_y)L$. From Eq. (2), we know that the transmission depends not only on the phase difference ($\varphi, 2\varphi$), but also on the polarization angle $\alpha$ of the input light and rotation angle $\theta$. When $a_2 = 0.5$, it is only a dual-pass M-Z interferometer. While $a_2 \neq 0.5$ (for this experiment, $a_2 = 0.3$), the proposed comb filter shows dual-function operation, the tunable channel spacing and the interleaving operation, by appropriately rotating the PC1 ($\alpha$) and the PC3 ($\theta$). For example, when $\theta = 0.65\pi$, the filter can act as a single-pass ($\alpha = 0.175\pi$) and a dual-pass ($\alpha = 0.425\pi$) M-Z interferometer. Furthermore, the wavelength switching (interleaving) operation can be achieved (i.e., for single-pass operation, $\alpha = 0.175\pi, 0.675\pi$). It is also important to note that the spacing tunable and interleaving operation are periodic processes with the periods of $\pi/4$ and $\pi/2$, respectively. Fig. 2 shows the measured tunable transmission spectra of the filter by rotating the PC1 when the PC3 was fixed in a proper orientation. As illustrated in Fig. 2, dual-function operation, the comb spacing tunable and the wavelength switching (interleaving) operation can be easily achieved. Therefore, it can be employed in multi-wavelength fiber lasers to realize tunable and switchable lasing operation.

In the experiment, the pump power was fixed at 70 mW. When the PZT was driven by a sinusoidal signal waveform with a frequency of 4.2 kHz, the laser operated in stable multi-wavelength lasing simultaneously. The PC3 was first rotated in a proper position where could realize the tunable operation and fixed it in the following experiment. Actually, the laser cavity incorporating a polarizer produces the wavelength-dependent polarization rotation mechanism, then the polarization states of different wavelengths are diversified, and each wavelength could have different loss. This contributes to optimize stable multi-wavelength output. Therefore, if the PCs were not properly set, the lasing lines were few and uneven. In the experimental observation, we found that the channel-spacing tunability is more sensitive to the PC1 setting while the number of the lasing lines and the lasing locations are more dependent on the PC2. It is because that the polarization state launched into the filter was mainly controlled by the PC1 and this determines the tunability of the filter according to the theoretical analysis. While the PC2 is more sensitive to adjust the polarization states of the lasing light that launched into the PD-ISO which facilitate the control of the wavelength-dependent polarization rotation effect. Fig. 3 shows the typical output spectra when the PC2 was not adjusted properly but only rotated the PC1. The channel spacing tunability between 0.4 nm and 0.8 nm and wavelength switching operation could be obtained, which was well consistent with the filter spectral spacing shown in Fig. 2. It is to note that this spacing tunable process is reversible. However, in this case the lasing lines were few and uneven.

When the PCs were further properly adjusted, the number of the lasing lines, the lasing flatness and the lasing locations could be improved. As the PCs were in optimum positions, up to 29 stable lasing lines with 0.4 nm spacing and 14 lasing wavelengths with 0.8 nm spacing were achieved. Fig. 4 presents the output spectra of a maximum number of 29 lasing wavelengths with 0.4 nm spacing in 3 dB bandwidth. Then we rotated the PC1 placed before the comb filter slightly, the evolution of wavelength spacing from 0.4 nm to 0.8 nm was clearly observed in the experiment. Finally, the stable 14 lasing wavelengths with a channel spacing of 0.8 nm in 3 dB bandwidth were obtained, as shown in Fig. 5. In these cases, the lasing wavelength switching operation was also observed. In order to verify the stability of the proposed multi-wavelength fiber laser, we repeatedly scanned the output of the proposed fiber laser in 30 minutes. Fig.6 shows the power variation of four individual channels and the wavelength drift in the experimental observation. The maximum power fluctuation is less than 1.0 dB and the maximum wavelength drift is 0.05 nm. These results indicate that the proposed fiber laser operated stably at room temperature.

In conclusion, we have proposed and demonstrated a tunable and switchable multi-wavelength EDF ring laser based on a new type comb filter. The comb filter provides the wavelength spacing tunability between 0.4 nm and 0.8 nm as well as the lasing wavelength switching operation. Correspondingly, up to 29 stable lasing lines with 0.4 nm spacing and 14 lasing lines with 0.8 nm spacing were obtained. Moreover, the lasing output, including the number of the lasing lines, the lasing evenness and the lasing locations, can also be flexibly adjusted.

Figure Captions

Fig. 1. The schematic of the proposed multi-wavelength fiber laser.

Fig. 2. The measured wavelength spacing tunable transmission spectra of the comb filter.

Fig. 3. The typical output spectra when the PC2 was not in proper setting.

Fig. 4. The stable output under the 29 lasing lines operation with a spacing 0.4 nm.

Fig. 5. The stable 14 lasing wavelengths with a channel spacing of 0.8 nm.

Fig. 6. The power variation of four individual channels and the wavelength drift in 30 minutes.

Fig. 1

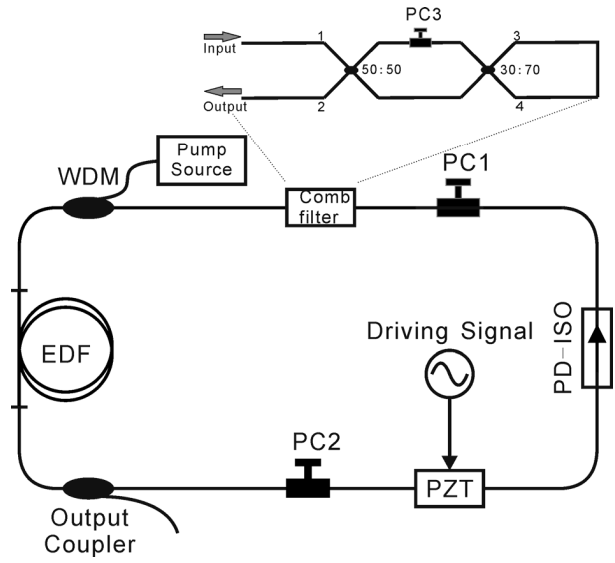

Fig. 1. The schematic of the proposed multi-wavelength fiber laser.

Fig. 2

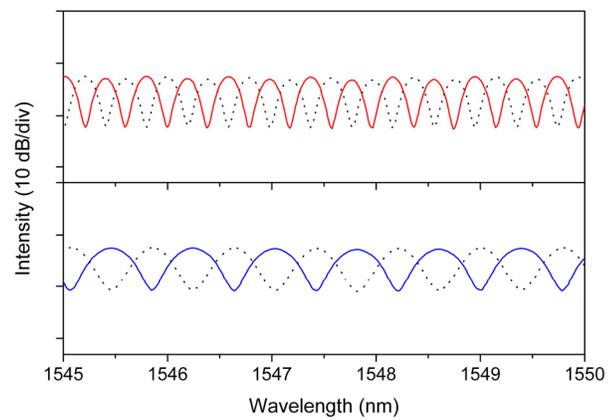

Fig. 2. The measured wavelength spacing tunable transmission spectra of the comb filter.

Fig. 3

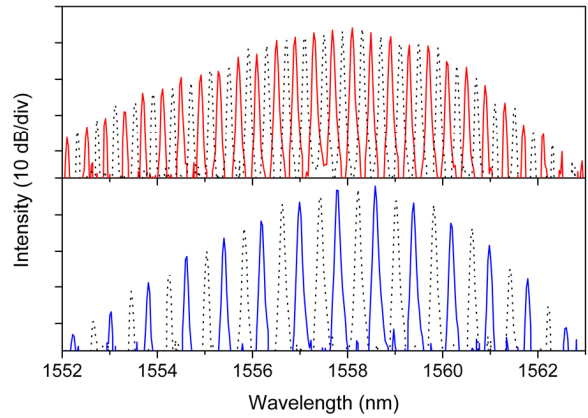

Fig. 3. The typical output spectra when the PC2 was not in proper setting.

Fig. 4

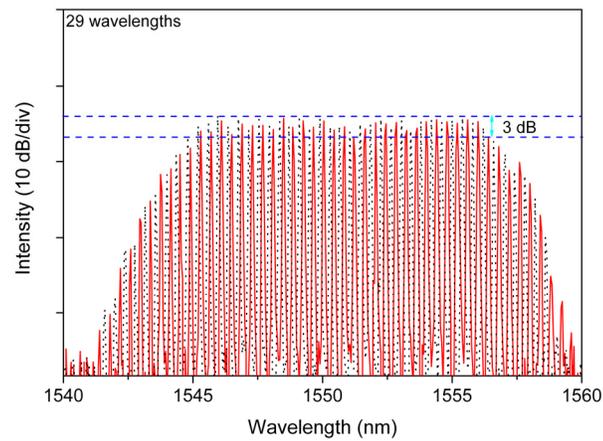

Fig. 4. The stable output under the 29 lasing lines operation with a spacing 0.4 nm.

Fig. 5

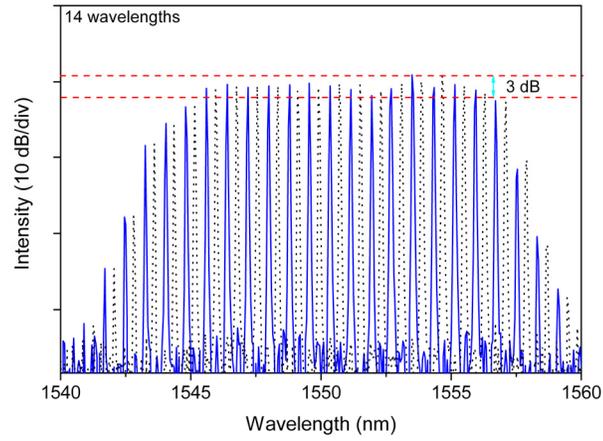

Fig. 5. The stable 14 lasing wavelengths with a channel spacing of 0.8 nm.

Fig. 6

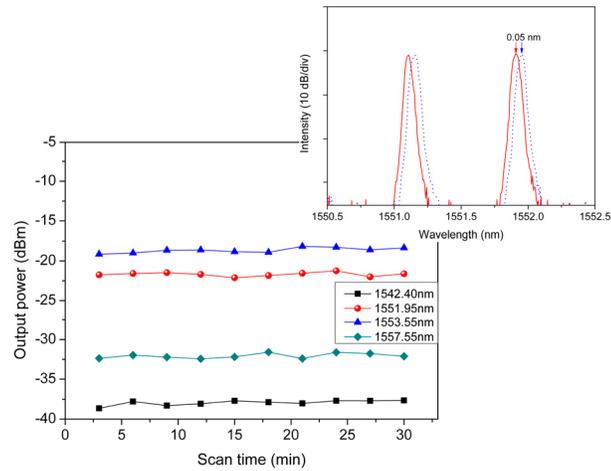

Fig. 6. The power variation of four individual channels and the wavelength drift in 30 minutes.